\documentclass{article}
\usepackage{spconf,amsmath,graphicx}
\usepackage[subtle]{savetrees}

\usepackage{amsmath}
\usepackage{amsfonts}  
\allowdisplaybreaks[4]
\usepackage[short]{optidef}

\usepackage{graphicx}
\usepackage{float}    
\usepackage{caption}
\usepackage{subcaption}
\usepackage{pdfpages}   
\usepackage{pdfpages}   
\usepackage{epstopdf}
\usepackage[short]{optidef}

\usepackage{cite}
\usepackage{url} 

\usepackage[ruled,vlined]{algorithm2e}  

\title{Waveform Optimization for Wireless Power Transfer with Power Amplifier and Energy Harvester Non-Linearities}
\name{Yumeng Zhang and {Bruno Clerckx}  }
\address{Department of Electrical and Electronic Engineering\\Imperial College London, London, UK}
\begin{document}
%
\maketitle
\begin{abstract}
Waveform optimization has recently been shown to be a key technique to boost the efficiency and range of far-field wireless power transfer (WPT). Current research has optimized transmit waveform adaptive to channel state information (CSI) and accounting for energy harvester (EH)’s non-linearity but under the assumption of linear high power amplifiers (HPA) at the transmitter. This paper proposes a channel-adaptive waveform design strategy that optimizes the transmitter's input waveform considering both HPA and EH non-linearities. Simulations demonstrate that HPA’s non-linearity degrades the energy harvesting efficiency of WPT significantly, while the performance loss can be compensated by using the proposed optimal input waveform.
\end{abstract}
\begin{keywords}
Waveform design, energy harvesting, non-linearities, power amplifier, wireless power transfer
\end{keywords}
\section{Introduction}
\label{sec:intro}
Far-field WPT is considered as a promising technique to exert a revolutionary impact on the powering systems of low power devices and to be the enabler of 1G mobile power networks \cite{BrunoToward}. Nevertheless, boosting the efficiency of WPT remains a key challenge \cite{clerckx2021wireless}. For this purpose, early efforts in the RF community have focused on the design of efficient rectennas \cite{suh2002high,1556784}, while recent efforts in the communication community have emphasized the crucial benefits of efficient signal designs for WPT \cite{Clerckx2016Waveform}. 

Of notable importance is the work in \cite{Clerckx2016Waveform} that developed a systematic framework for the design and optimization of waveforms to maximize the harvested DC power at the output of the rectenna. Such waveform optimization was further extended to other scenarios such as limited-feedback\cite{Huang1}, large-scale \cite{HuangLarge2017}, multi-user \cite{HuangLarge2017,abeywickrama2021refined}, opportunistic/fair-scheduling \cite{kim2020opportunistic,8476162}, multi-input-multi-output \cite{shen2020beamforming}, low-complexity \cite{ClerckxA}, prototyping and experimentation \cite{KimSignal}, wireless information and power transfer (WIPT) \cite{clerckx2017wireless} and wireless powered backscatter communications \cite{clerckx2017wirelessly}. 

Despite those progress, the above waveform optimization was performed without much consideration for HPA's non-linearity at the transmitter. Indeed, it has been verified that HPA's non-linearity distorts the amplitude and phase of its input signal\cite{santella1998hybrid}, and results in unexpected performance degradation particularly with multi-sine waveform transmission where the amplitudes' high variations make the input signal more vulnerable to HPA's non-linearity \cite{park2020performance}.

To combat HPA's non-linear effect, mainly two lines of methods have been put forward, namely designing signals less susceptible to HPA's non-linearity and by means of digital pre-distortion (DPD). The former method decreases input signals' exposure to HPA's non-linear region by limiting their amplitude variations, such as peak-to-average-power-ratio (PAPR) reduction \cite{kryszkiewicz2018amplifier}, distortion power reduction across desired bandwidth \cite{kryszkiewicz2018amplifier} and leakage power reduction across adjacent channel\cite{goutay2021end}. Indeed, PAPR reduction has been introduced as a transmit waveform constraint in WPT in \cite{Clerckx2016Waveform}. However, this class of methods might be less efficient in WPT because HPA's power efficiency is often higher in the non-linear region and also because the method is not adaptive to HPAs' characteristics. In contrast, DPD pre-distorts the desired input signal according to HPA's transfer characteristics to linearize the transfer function of the joint pre-distorter-and-HPA structure \cite{fu2014frequency}. Recent literature has revealed the performance gain of using DPD in simultaneous WIPT (SWIPT) systems, observing an improved rate-energy region \cite{2020WIPTNON}. However, those papers did not propose a waveform design strategy that comprises HPA's non-linearity and EH's non-linearity simultaneously in WPT/SWIPT\cite{krikidis2020information}.

This letter proposes a practical WPT system model accounting for both HPA and rectenna non-linearity, and derives the optimal waveform solution in the non-linear system based on a non-linear solid-state power amplifier (SSPA) and the non-linear rectenna in \cite{Clerckx2016Waveform}. Simulations verify the benefit of the proposed waveform, which compensates the power loss caused by HPA's non-linearity. The paper is organised as follows. Section \ref{section_WPT_system_model} models the non-linear WPT architecture. Section \ref{section_optimization} declares the optimization problem and reformulates it into a tractable problem, which is solved by successive convex programming (SCP), combining with Barrier's method and the gradient descent (GD) method. Section \ref{section_simulations} presents simulation results, and Section \ref{section_conclusion} draws the conclusions.
\begin{figure}[htb]

\begin{minipage}[b]{1.0\linewidth}
  \centering
  \centerline{\includegraphics[width=9cm]{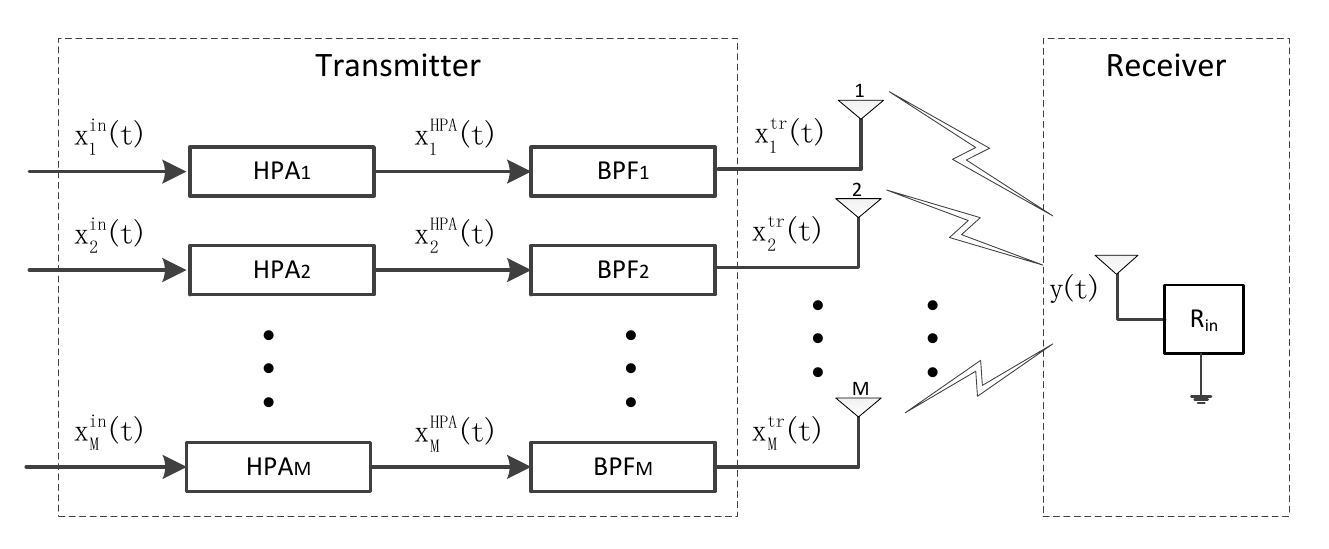}}
\end{minipage}
\caption{The WPT structure with HPA and rectenna non-linearity.}
\label{Fig_whole_structure}
\end{figure}

\section{WPT System Model}
\label{section_WPT_system_model}
Consider a system as depicted in Fig. \ref{Fig_whole_structure}. The transmitter consists of $M$ antennas, with each antenna transmitting over $N$ evenly frequency-spaced sub-carriers. At the transmitter, the RF signal is amplified and filtered before being transmitted. The complex input signal at the amplifier of the $m^{\text{th}}\:\:(m=1,2,...,M)$ antenna is written as:
\begin{align}
\label{eq_input_signal_complex}
\widetilde{x}^{\text{in}}_m(t)&=\sum_{n=0}^{N-1}\widetilde{w}^{\text{in}}_{n,m}e^{j2\pi f_nt},
\end{align}
where $\widetilde{w}^{\text{in}}_{n,m}$ denotes the complex weight of the $n^{\text{th}}\:\:(n=0,1,...,N-1)$ sub-carrier at the $m^{\text{th}}$ antenna, and $f_n=f_0+(n-1)\Delta_f$ denotes the frequency of the $n^{\text{th}}$ sub-carrier, with $f_0$ being the lowest sub-carrier frequency and $\Delta_f$ being the frequency spacing.

The input signal $\widetilde{x}^{\text{in}}_m(t)$ is amplified and filtered before being transmitted. Adopting an SSPA model\cite{rapp1991effects}, the complex signal at the output of the SSPA at the $m^{\text{th}}$ antenna becomes:
\begin{equation}
\label{eq_HPA_model}
\widetilde{x}^{\text{HPA}}_m(t)=f_{\text{SSPA}}(\widetilde{x}^{\text{in}}_m(t))=\frac{G\widetilde{x}^{\text{in}}_m(t)}{[1+(\frac{Gx^{\text{in}}_m(t)}{A_s})^{2\beta}]^{\frac{1}{2\beta}}},
\end{equation}
where $x^{\text{in}}_m(t)=|\widetilde{x}^{\text{in}}_m(t)|$ is the amplitude envelop of the complex input signal $\widetilde{x}^{\text{in}}_m(t)$, $G$ denotes the small-signal amplifier gain of SSPA, $A_s$ denotes the saturation voltage of SSPA, and $\beta$ denotes the smoothing parameter of SSPA.

$\widetilde{x}^{\text{HPA}}_m(t)$, after propagating through a BPF, becomes the complex transmit signal $\widetilde{x}^{\text{tr}}_m(t)$. Denote by $\widetilde{w}^{\text{tr}}_{n,m}$ the complex weight of the $n^{\text{th}}$ sub-carrier at the $m^{\text{th}}$ antenna. We have:
\begin{align}
\label{eq_transmit_signal_complex}
\widetilde{x}^{\text{tr}}_m(t)&=\sum_{n=0}^{N-1}\widetilde{w}^{\text{tr}}_{n,m}e^{j2\pi f_nt}.
\end{align}

After propagating through the frequency-selective channel, the complex received signal at the receiver is:
\begin{align}
\label{eq_WPT_received_signal}
\widetilde{y}(t)&=\sum_{m=1}^{M}\sum_{n=0}^{N-1}\widetilde{h}_{n,m}\widetilde{w}^{\text{tr}}_{n,m}e^{j2\pi f_nt},
\end{align}
where $\widetilde{h}_{n,m}\sim \mathcal{CN}(0,1)$ denotes the complex channel of the $n^{\text{th}}$ sub-carrier of the signal from the $m^{\text{th}}$ transmit antenna.

At the receiver, the wireless signal $\widetilde{y}(t)$ is picked up and is converted into DC as a power supply via a rectenna.  We model the non-linear rectenna based on \cite{Clerckx2016Waveform}, whose output DC is approximately proportional to a scaling term as:
\begin{align}
\label{eq_scaling_term0}
z_{DC}&=k_2R_{\text{ant}}\varepsilon\{y(t)^2\}+k_4R_{\text{ant}}^2\varepsilon\{y(t)^4\}\\
\label{eq_SSPA_poly_x_tr}
\nonumber&=\frac{k_2R_{\text{ant}}}{2}(\sum_{m=1}^{M}\sum_{n=0}^{N-1}|\widetilde{w}^{\text{tr}}_{n,m}\widetilde{h}_{n,m}|^2)\\
\nonumber&\quad +\frac{3k_4R_{\text{ant}}^2}{8}(\sum_{\tiny{\begin{array}{c}m_0,m_1\\m_2,m_3\end{array}}}\sum_{\tiny{\begin{array}{c} n_0,n_1,n_2,n_3\\n_0+n_1=n_2+n_3\end{array}}} \widetilde{h}_{n_0,m_0}\widetilde{w}^{\text{tr}}_{n_0,m_0}\times\\
&\:\:\:\:\:\:\:\:\:\:\:\:\:\:\:\:\:\:\:\:\widetilde{h}_{n_1,m_1}\widetilde{w}^{\text{tr}}_{n_1,m_1}\widetilde{h}^*_{n_2,m_2}\widetilde{w}^{\text{tr}^*}_{n_2,m_2}\widetilde{h}^*_{n_3,m_3}\widetilde{w}^{\text{tr}^*}_{n_3,m_3}),
\end{align}
where $y(t)=\mathfrak{R}\{y(t)\}$ is the real received signal, and $k_i=i_s/(i!(\eta_0 V_0)^i)$ with $i_s$ being the reverse bias saturation current, $\eta_0$ being the ideality factor, $V_0$ being the thermal voltage of the diode and $R_{\text{ant}}$ being the characteristic impedance of the receiving antenna.

\section{Optimization Solutions}
\label{section_optimization}
Consequently, subjected to a transmit power constraint and an input power constraint,  the optimization problem to maximize the end-to-end harvested DC in WPT is written as:
\begin{maxi!}
        {{\{\widetilde{w}^{\text{in}}_{n,m}\}}}{z_{DC}(\{\widetilde{w}^{\text{in}}_{n,m}\}),}{\label{eq_optimization_P1}}{\label{eq_optimization_P1_1}}
        \addConstraint{\frac{1}{2}\sum_m^M \sum_n^N |\widetilde{w}^{\text{in}}_{n,m}|^2 \leq P^{\max}_{\text{in}}}\label{eq_optimization_P1_2}
        \addConstraint{\frac{1}{2}\sum_m^M \sum_n^N |\widetilde{w}^{\text{tr}}_{n,m}(\{\widetilde{w}^{\text{in}}_{n,m}\})|^2  \leq P^{\max}_{\text{tr}},}\label{eq_optimization_P1_3}
      \end{maxi!}
where $P^{\max}_{\text{in}}$ and $P^{\max}_{\text{tr}}$ are the input power constraint and the transmit power constraint respectively \footnote{ Eq. \eqref{eq_optimization_P1_2} prevents the power of SSPA's input signal exceeding SSPA's saturation power (the maximal output power) significantly, and avoids poor amplifier efficiency. Eq. \eqref{eq_optimization_P1_3} limits the transmit signal's RF exposure to human beings.}.

Unfortunately, the scaling term $z_{DC}$ as a function of $\{\widetilde{w}^{\text{in}}_{n,m}\}$ in Eq. \eqref{eq_optimization_P1_1} is hardly specified, while $z_{DC}$ as a function of $\{\widetilde{w}^{\text{tr}}_{n,m}\}$ has been written explicitly in Eq. \eqref{eq_SSPA_poly_x_tr}. Thus, to solve problem \eqref{eq_optimization_P1}, we alter the optimization variables in problem \eqref{eq_optimization_P1} from $\{\widetilde{w}^{\text{in}}_{n,m}\}$ into $\{\widetilde{w}^{\text{tr}}_{n,m}\}$ and express $\{\widetilde{w}^{\text{in}}_{n,m}\}$ in Eq. \eqref{eq_optimization_P1_2}  by using $\{\widetilde{w}^{\text{tr}}_{n,m}\}$. Consequently, an equivalent optimization problem is formed as:
      \begin{maxi!}
        {\substack{\{\overline{w}^{\text{tr}}_{n,m}\},\{\widehat{w}^{\text{tr}}_{n,m}\}}}{z_{DC}(\{\overline{w}^{\text{tr}}_{n,m}\},\{\widehat{w}^{\text{tr}}_{n,m}\}),}{\label{eq_optimization_P3}}{\label{eq_optimization_P3_1}}
        \addConstraint{{\sum_{m=1}^M\frac{1}{2T}\int_{T}\{\frac{x^{\text{tr}}_m(t)}{G}[\frac{1}{1-(\frac{x^{\text{tr}}_m(t)}{A_s})^{2\beta}}]^{\frac{1}{2\beta}}\}^2 dt}\nonumber\breakObjective{\leq P^{\max}_{\text{in}}}}\label{eq_optimization_P3_3}
       \addConstraint{\frac{1}{2}\sum_m^M \sum_n^N {\overline{w}^{\text{tr}^2}_{n,m}}+\widehat{w}^{\text{tr}^2}_{n,m} \leq P^{\max}_{\text{tr}},}\label{eq_optimization_P3_2}
      \end{maxi!}
where $\{\overline{w}^{\text{tr}}_{n,m}\}$ and $\{\widehat{w}^{\text{tr}}_{n,m}\}$ are the real and imaginary part of $\{\widetilde{w}^{\text{tr}}_{n,m}\}$ respectively, and $x^{\text{tr}}_m(t)$ in Eq. \eqref{eq_optimization_P3_3} is the amplitude of $\widetilde{x}^{\text{tr}}_m(\{\overline{w}^{\text{tr}}_{n,m}\},\{\widehat{w}^{\text{tr}}_{n,m}\},t)$.  The objective function and constraints in problem \eqref{eq_optimization_P3} can be proved convex.

Problem \eqref{eq_optimization_P3} maximizes a convex objective function, which can be solved by SCP. In SCP, the objective term is linearly approximated by its first-order Taylor expansion at a fixed operating point, forming a new tractable optimization problem whose optimal solution is used as a new operating point of the next iteration. The procedure is repeated until two successive solutions are close enough and can be viewed as the solution of problem \eqref{eq_optimization_P3}. Assume $(\{\overline{w}^{\text{tr},(l-1)}_{n,m}\},\{\widehat{w}^{\text{tr},(l-1)}_{n,m}\})$ are the values of the operating point at the beginning of the $l^{\text{th}}$ iteration. Then, $z_{DC}(\{\overline{w}^{\text{tr}}_{n,m}\},\{\widehat{w}^{\text{tr}}_{n,m}\})$ at the $l^{\text{th}}$ iteration is linearly approximated as:
\begin{align}
\label{eq_first_order_Taylor}
z_{DC}^{(l)}(\{\overline{w}^{\text{tr}}_{n,m}\},\{\widehat{w}^{\text{tr}}_{n,m}\})=\sum_{m=1}^M\sum_{n=0}^{N-1} \overline{\alpha}^{(l)}_{n,m}\overline{w}^{\text{tr}}_{n,m}+\widehat{\alpha}^{(l)}_{n,m}\widehat{w}^{\text{tr}}_{n,m},
\end{align}
where $(\{\overline{\alpha}^{(l)}_{n,m}\},\{\widehat{\alpha}^{(l)}_{n,m}\})$ are the first-order Taylor coefficients of $(\{\overline{w}^{\text{tr}}_{n,m}\},\{\widehat{w}^{\text{tr}}_{n,m}\})$ respectively at the $l^{\text{th}}$ iteration.

Hence, at the $l^{\text{th}}$ iteration, problem \eqref{eq_optimization_P3} is approximated as:
\begin{maxi!}
        {\substack{\{\overline{w}^{\text{tr}}_{n,m}\},\{\widehat{w}^{\text{tr}}_{n,m}\}}}{z_{DC}^{(l)}(\{\overline{w}^{\text{tr}}_{n,m}\},\{\widehat{w}^{\text{tr}}_{n,m}\}),}{\label{eq_optimization_P4}}{\label{eq_optimization_P4_1}}
        \addConstraint{\text{Eq}. \eqref{eq_optimization_P3_2},\quad \text{Eq}. \eqref{eq_optimization_P3_3}.}{\label{eq_optimization_P4_2}}
      \end{maxi!}

Problem \eqref{eq_optimization_P4} is solved by using Barrier's method, where the non-linear constraints in Eq. \eqref{eq_optimization_P4_2} are omitted by reformulating problem \eqref{eq_optimization_P4} into:
\begin{align}
\label{eq_optimization_P4_l}
\nonumber\min_{\{\overline{w}^{tr}_{n}\},\{\widehat{w}^{tr}_{n}\}} \quad  &-z_{DC}^{(l)}(\{\overline{w}^{\text{tr}}_{n,m}\},\{\widehat{w}^{\text{tr}}_{n,m}\})\\
&\quad +\sum_{i=1}^{2}I_-(f_{c,i}(\{\overline{w}^{\text{tr}}_{n}\},\{\widehat{w}^{\text{tr}}_{n}\})),
\end{align}
where 
\begin{align}
\label{eq_interpratation}
I_-(x)=&\lbrace\begin{matrix}
0,\:\:\:\:&x\leq 0,\\
\infty,\:\:\:\:&x> 0,
\end{matrix}\\\label{eq_interpratation1}
f_{c,1}(\{\overline{w}^{\text{tr}}_{n}\},\{\widehat{w}^{\text{tr}}_{n}\})=&\frac{1}{2}\sum_m^M \sum_n^N {\overline{w}^{\text{tr}^2}_{n,m}}+\widehat{w}^{\text{tr}^2}_{n,m} - P^{\max}_{\text{tr}},\\
\nonumber f_{c,2}(\{\overline{w}^{\text{tr}}_{n}\},\{\widehat{w}^{\text{tr}}_{n}\})=&\sum_{m=1}^M\frac{1}{2T}\int_{T}\{\frac{x^{\text{tr}}_m(t)}{G}[\frac{1}{1-(\frac{x^{\text{tr}}_m(t)}{A_s})^{2\beta}}]^{\frac{1}{2\beta}}\}^2 dt\\
& - P^{\max}_{\text{in}}.
\end{align}

Further, to make problem \eqref{eq_optimization_P4_l} differentiable, $I_-(x)$ is approximated as:
\begin{equation}
\label{eq_I_-}
\widehat{I}_-(x)=-(\frac{1}{t})\log(-x),
\end{equation}
where $t$ is a parameter that sets the accuracy of the approximation. The larger the $t$, the closer the $\widehat{I}_-(x)$ is to ${I}_-(x)$.

Consequently, for a specific $t$, the optimization problem \eqref{eq_optimization_P4_l}  becomes:
\begin{align}
\label{eq_optimization_barrier_approx}
\nonumber \min_{\{\overline{w}^{\text{tr}}_{n}\},\{\widehat{w}^{\text{tr}}_{n}\}} \quad  &-z_{DC}^{(l)}(\{\overline{w}^{\text{tr}}_{n,m}\},\{\widehat{w}^{\text{tr}}_{n,m}\})-\\&\frac{1}{t}\sum_{i=1}^{2}\log(-f_{c,i}(\{\overline{w}^{\text{tr}}_{n}\},\{\widehat{w}^{\text{tr}}_{n}\})),
\end{align}
which can be solved by GD methods such as Newton's Method.

In summary, the optimization problem \eqref{eq_optimization_P3} is solved in a iterative manner by adopting SCP. In each SCP's round, the corresponding optimization problem \eqref{eq_optimization_P4} is solved by Barrier's method iteratively, with an exit condition of a sufficient large $t$ so that problem  \eqref{eq_optimization_barrier_approx} approximates problem \eqref{eq_optimization_P4} satisfyingly. The whole optimization process is described in Algorithm \ref{SCP}.
\begin{algorithm}[h]
\SetAlgoLined
 $\textbf{Input}$: $(\{\overline{w}^{\text{tr}}_{n}\},\{\widehat{w}^{\text{tr}}_{n}\})^{(0)},\epsilon_0>0,l\leftarrow 1$\;
 $\textbf{Output}$: $(\{\overline{w}^{\text{tr}}_{n}\},\{\widehat{w}^{\text{tr}}_{n}\})^{\star}$\;
 $\textbf{Repeat}$: \\
$\:\:\:\:\:\:1: \:$Compute $(\{\overline{\alpha}\},\{\widehat{\alpha}\})^{(l)}$ at the operating point $(\{\overline{w}^{\text{tr}}_{n}\},\{\widehat{w}^{\text{tr}}_{n}\})^{(l-1)}$ using Taylor expansion\;
$\:\:\:\:\:\:2: \text{Compute } (\{\overline{w}^{\text{tr}}_{n}\},\{\widehat{w}^{\text{tr}}_{n}\})^{(l)}$ using Algorithm \ref{algorithm_barrier}\; 
$\:\:\:\:\:\:3: \:$Update $(\{\overline{w}^{\text{tr}}_{n}\},\{\widehat{w}^{\text{tr}}_{n}\})^{\star}\leftarrow (\{\overline{w}^{\text{tr}}_{n}\},\{\widehat{w}^{\text{tr}}_{n}\})^{(l)}$\;
$\:\:\:\:\:\:4: \:$Quit if \\
$\:\:\:\:\:\:\:\:\:\:\:\:\:|{(\{\mathbf{\overline{w}}^{\text{tr}}_{n}\},\{\mathbf{\widehat{w}}^{\text{tr}}_{n}\})^{(l)}}-{(\{\mathbf{\overline{w}}^{\text{tr}}_{n}\},\{\mathbf{\widehat{w}}^{\text{tr}}_{n}\})^{(l-1)}}|< \epsilon_0$\;
$\:\:\:\:\:\:\:5: \:l\leftarrow l+1$\;
\caption{Successive convex programming (SCP)}
\label{SCP}
\end{algorithm}

\begin{algorithm}[h]
\SetAlgoLined
 $\textbf{Input}$: $(\{\overline{w}^{\text{tr}}_{n}\},\{\widehat{w}^{\text{tr}}_{n}\})^{(B_0)}\leftarrow(\{\overline{w}^{\text{tr}}_{n}\},\{\widehat{w}^{\text{tr}}_{n}\})^{(l-1)},\:t>0,$\\
 $\:\:\:\:\:\:\:\:\:\:\:\:\:\:\:\:\:\:\:\:\:\:\:\:\:\:\:\:\:\:\:\:\:\:\:\:\:\:\:\:\:\:\:\:\:\:\:\:\:\:\:\:\mu_B>0,\epsilon_B>0$\;
 $\textbf{Output}$: $(\{\overline{w}^{\text{tr}}_{n}\},\{\widehat{w}^{\text{tr}}_{n}\})^{(l)}$; \\
 $\textbf{Repeat}$: \\
 $\:\:\:\:\:\:\:1:\:$Compute $(\{\overline{w}^{\text{tr}}_{n}\},\{\widehat{w}^{\text{tr}}_{n}\})$ by minimizing problem \eqref{eq_optimization_barrier_approx} using Newton's Method with initialised point $(\{\overline{w}^{\text{tr}}_{n}\},\{\widehat{w}^{\text{tr}}_{n}\})^{(B_0)}$\;
 $\:\:\:\:\:\:\:2:\text{Update }(\{\overline{w}^{\text{tr}}_{n}\},\{\widehat{w}^{\text{tr}}_{n}\})^{(l)}\leftarrow(\{\overline{w}^{\text{tr}}_{n}\},\{\widehat{w}^{\text{tr}}_{n}\})$\;
 $\:\:\:\:\:\:\:3: \:\text{Quit if } 2/t < \epsilon_B$\;
$\:\:\:\:\:\:\:4: \:t\leftarrow\mu_Bt,\:(\{\overline{w}^{\text{tr}}_{n}\},\{\widehat{w}^{\text{tr}}_{n}\})^{(B_0)}\leftarrow(\{\overline{w}^{\text{tr}}_{n}\},\{\widehat{w}^{\text{tr}}_{n}\})^{(l)}$\;
\caption{Barrier's method}
\label{algorithm_barrier}
\end{algorithm}

\textit{Remark 1:} Current literature optimizes the WPT transmit waveform based on different optimization variables, such as the amplitude and phase of the weights\cite{2017Communications,shen2020beamforming}, the real and imaginary part of the weights\cite{abeywickrama2021refined}, and the complex weight vector\cite{HuangLarge2017}. This letter solves problem \eqref{eq_optimization_P3} by optimizing the real and imaginary part of the weights, because the non-linear SSPA constraint in Eq. \eqref{eq_optimization_P3_3} is only proved convex relative to the real and imaginary parts of the weights of the sub-carriers.

\section{Simulations}
\label{section_simulations}
The power efficiency of the proposed waveform is evaluated under a Wi-Fi-like scenario with $f_0=5.18$ GHz. For the SSPA, set the smoothing parameter to $\beta=1$ and the small-signal gain to $G=1$; For the rectenna, set $i_s=5\:\mu$A, $\eta_0=1.05$, $V_0=25.86$ mV, and $R_{\text{ant}}=50\:\Omega$. 

\begin{figure}[t]
\begin{minipage}[b]{1.0\linewidth}
  \centering
  \centerline{\includegraphics[width=8.5cm]{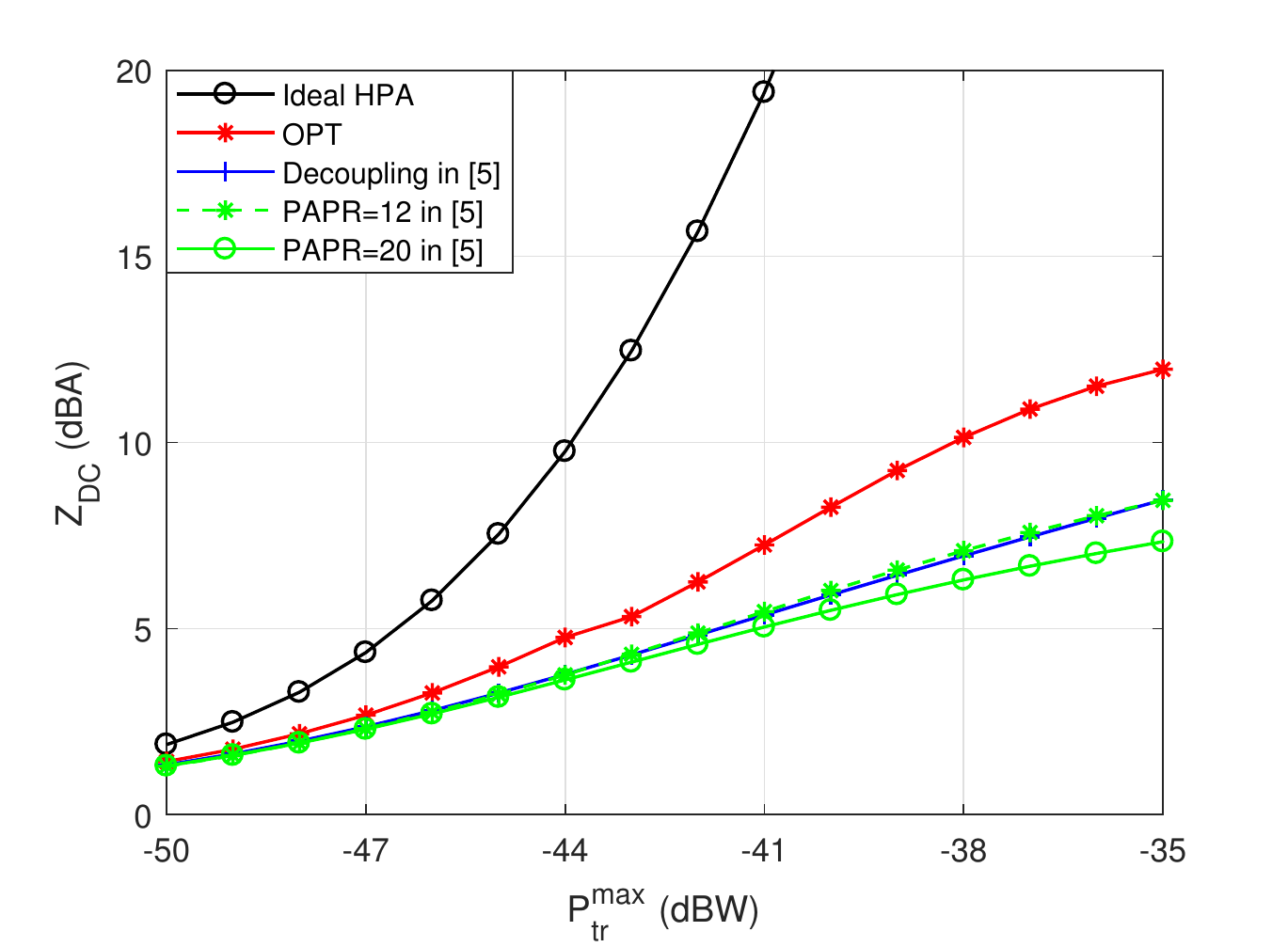}}
\end{minipage}
\caption{{Energy harvesting performance with $G=1, A_s=-35 \:$dBV,$\: P^{\max}_{\text{in}}=-20\:$dBW,$\: N=8$. `Ideal HPA' stands for using the optimal transmit waveform in \cite{Clerckx2016Waveform} to the input of an ideal HPA; `OPT' stands for the proposed optimal solution accounting for SSPA's and rectenna's non-linearity; `Decoupling' stands for using the optimal transmit waveform in \cite{Clerckx2016Waveform} to the input of SSPA; `PAPR=12' and `PAPR=20' stand for the optimal transmit waveform in \cite{Clerckx2016Waveform} with different PAPR constraints.}}
\label{fig_diff_P_tr}
\end{figure}

Fig. \ref{fig_diff_P_tr} compares the energy harvesting performance between the proposed input waveform and the waveform considering only rectenna's non-linearity by putting the optimal transmit waveform in\cite{Clerckx2016Waveform} directly into SSPA. The energy harvesting performance assuming an ideal linear HPA is plotted as a benchmark (black), demonstrating the power loss caused by HPA's non-linearity compared with other curves. The comparison with using an ideal HPA also reveals that,  although larger transmit power gives larger harvested energy in practical WPT systems, it also leads to more severe power loss caused by HPA's non-linearity. When the transmit power constraint grows sufficiently large, the harvested energy is limited by the saturation power of SSPA. 

Fig.\ref{fig_diff_P_tr} also verifies that, until the transmit power constraint reaches SSPA's saturation power ($-35\:$dBW), the proposed waveform always outperforms all the other solutions which are only optimized for rectenna's non-linearity. The result highlights the significance of considering HPA's non-linearity for waveform design. Interestingly, Fig.\ref{fig_diff_P_tr} also shows that, although the non-linear HPA prefers low-PAPR input signals, using the transmit waveform with PAPR constraints in \cite{Clerckx2016Waveform} as the input waveform (PAPR=$20$) will not necessarily outperform using the transmit waveform without PAPR constraints in \cite{Clerckx2016Waveform} as the input waveform. This might originate from a trade-off between the HPA non-linearity and the rectenna non-linearity, since high PAPR signals are preferred by rectenna's non-linearity, which is the opposite for SSPA \cite{Clerckx2016Waveform}. The phenomenon indicates that adding PAPR constraint only is not sufficient to grasp the HPA's non-linearity for optimal input waveform design and thus highlights the significance of designing waveforms adaptive to SSPA's transfer characteristics. However, that the curve of PAPR$=12$ outperforms the curve of PAPR$=20$ still illustrates SSPA's preference on low-PAPR signals.


\begin{figure}[t]
\begin{minipage}[b]{1.0\linewidth}
  \centering
  \centerline{\includegraphics[width=8.5cm]{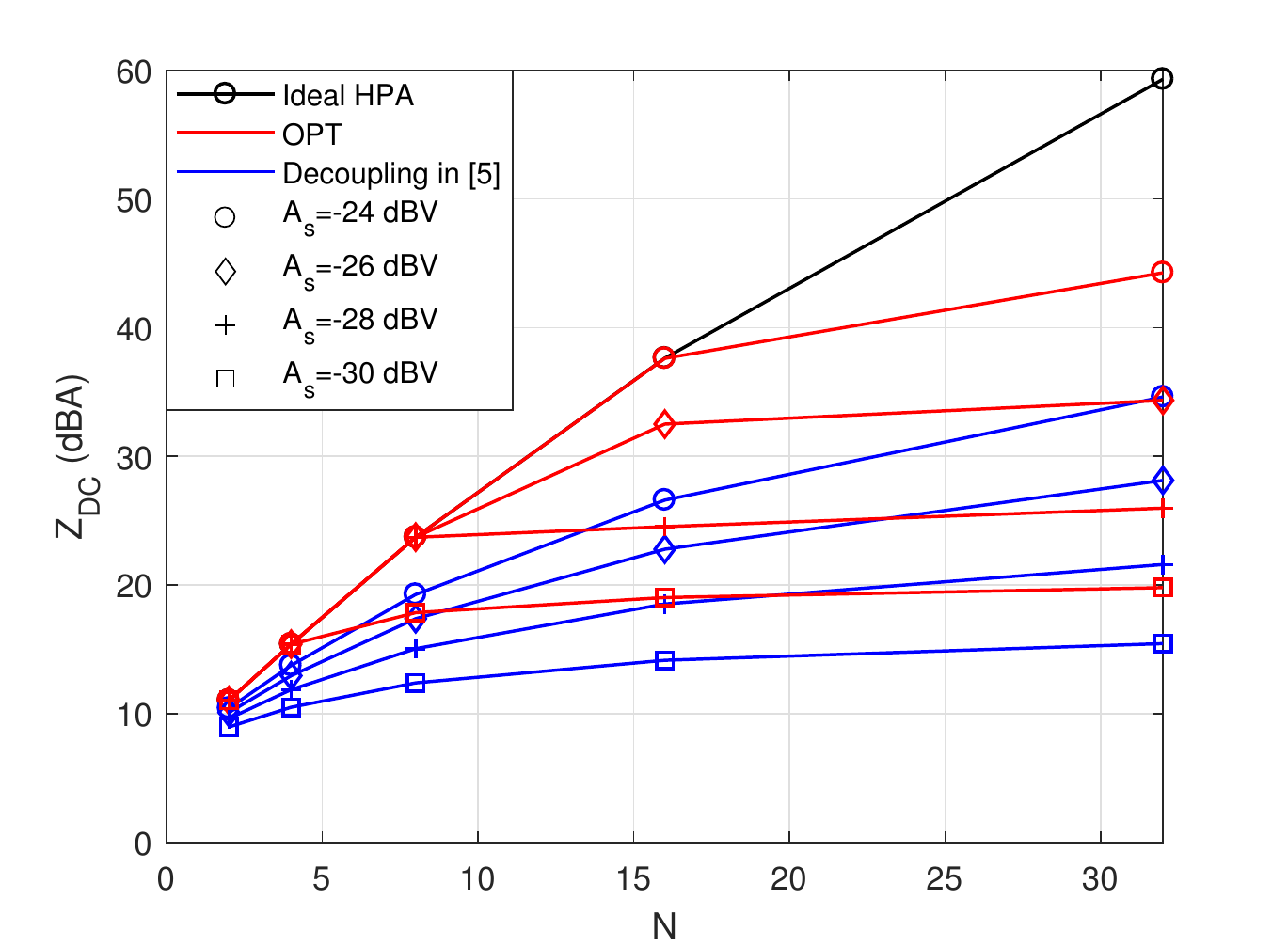}}
\end{minipage}
\caption{{$z_{DC}$ as a function of $N$ with different $A_s$, $G=1, \: P^{\max}_{\text{in}}=-20\:$dBW,  $P^{\max}_{\text{tr}}=-40\:$dBW.}}
\label{fig_diff_N}
\end{figure}

The effect of HPA's non-linearity on energy harvesting performance is further verified in Fig. \ref{fig_diff_N}, where $z_{DC}$ is plotted as a function of the number of the sub-carriers with different saturation voltages.  Fig. \ref{fig_diff_N} shows that the harvested energy increases linearly with the number of sub-carriers when using the optimal transmit waveform in \cite{Clerckx2016Waveform} into an ideal amplifier (black). However, if adopting the same waveform  as the black curve but with a non-linear SSPA (blue), the harvested energy tends to saturate when the number of sub-carriers keeps increasing, especially with low SSPA's saturation voltage. This is because the PAPR of the optimal waveform in \cite{Clerckx2016Waveform} increases with the number of sub-carriers, giving larger maximal amplitudes of the signal and making the signal exposed to SSPA's non-linear regime more severely, which results in more power loss. In contrast, using the proposed input waveform (red) can compensate for SSPA's non-linear effect and guarantee the same harvested energy as using an ideal amplifier, as long as the input signal does not make the SSPA operate in very high non-linear regime (i.e. $A_s=-24\:$dBV, $N=16$).


\section{Conclusions}
\label{section_conclusion}
This paper proposes an input waveform design strategy which maximizes the harvested energy in WPT, considering both HPA and rectenna non-linearity. The power loss caused by HPA's non-linearity is evaluated through simulations. The simulations also verify that the proposed input waveform achieves better energy harvesting performance compared with the waveform that only accounts for rectenna's non-linearity, emphasizing the significance of considering transmitter's non-linearity in efficient wireless powered networks design.

\bibliographystyle{IEEEtran}
\bibliography{references}

\end{document}